\documentclass[aps,nofootinbib,twocolumn]{revtex4}
\usepackage{graphicx,amsmath,hyperref,amssymb,bbm}
\newtheorem{theorem}{Theorem}
\newcommand{\slcm}{SL(2,\mathbbm {C})}
\newcommand{\slc}{\mbox{$\slcm$}}
\newcommand{\be}{\begin{equation}}
\newcommand{\ee}{\end{equation}}
\def\C{{\mathbbm C}}
\def\N{{\mathbbm N}}
\def\R{{\mathbbm R}}

\begin{document}
\title{ \Large Lorentz covariance of loop quantum gravity}
     \author{Carlo Rovelli}
     \affiliation{Centre de Physique Th\'eorique de Luminy\footnote{Unit\'e mixte de recherche du CNRS et des Universit\'es de Aix-Marseille I, Aix-Marseille II et Toulon-Var; affili\'e \`a la FRUMAM.}, Case 907, F-13288 Marseille, EU}
     \author{Simone Speziale}
     \affiliation{Centre de Physique Th\'eorique de Luminy\footnote{Unit\'e mixte de recherche du CNRS et des Universit\'es de Aix-Marseille I, Aix-Marseille II et Toulon-Var; affili\'e \`a la FRUMAM.}, Case 907, F-13288 Marseille, EU}
\date{\small  \today}
\begin{abstract}

\noindent 
The kinematics of loop gravity can be given a manifestly Lorentz-covariant formulation:  the conventional $SU(2)$-spin-network  Hilbert space can be mapped to a space $\cal K$ of $SL(2,\C)$ functions, where Lorentz covariance is manifest.  $\cal K$ can be described in terms of a certain subset of the ``projected" spin networks studied by Livine, Alexandrov and Dupuis. It is formed by $SL(2,\C)$ functions completely determined by their restriction on $SU(2)$. These are square-integrable in the $SU(2)$  scalar product, but not in the $SL(2,\C)$ one. Thus, $SU(2)$-spin-network states can be represented by Lorentz-covariant $SL(2,\C)$ functions, as two-component photons can be described in the Lorentz-covariant Gupta-Bleuler formalism.  As shown by Wolfgang Wieland in a related paper, this manifestly Lorentz-covariant formulation can also be directly obtained from canonical quantization.   We show that the spinfoam dynamics of loop quantum gravity is locally $SL(2,\C)$-invariant in the bulk, and yields states that are preciseley in $\cal K$ on the boundary.  This clarifies how the $SL(2,\C)$ spinfoam formalism yields an $SU(2)$ theory on the boundary.  These structures define a tidy Lorentz-covariant formalism for loop gravity. 
\end{abstract}

\maketitle

\section{Introduction}

General relativity (GR) has a local Lorentz symmetry.  Here we discuss the Lorentz covariance of loop quantum gravity (LQG)  in the spinfoam and canonical formalisms.  

The state space ${\cal H}_{SU(2)}$ of canonical LQG is defined in a fixed gauge, thus manifest local Lorentz covariance is broken. The lack of manifest Lorentz covariance has been often pointed out as an unpalatable feature of canonical LQG. Is LQG consistent with the local Lorentz invariance of GR? Can we reformulate the LQG kinematics in a manifestly Lorentz covariant language?  The ``projected spin network" formalism and recent developments in spinfoam theory bring light to this question.

A spinfoam definition of the LQG dynamics has been fast developing in the last few years \cite{Engle:2007uq,Livine:2007vk,Engle:2007qf,Engle:2007wy,Livine:2007ya,Freidel:2007py,Kaminski:2009fm} and is summarized in \cite{Rovelli:2010vv}.  The theory is built in an $SL(2,\C)$-covariant formalism and determines transition amplitudes between boundary states.  Here we observe that the boundary states of the spinfoam theory can be represented as functions on $SL(2,\C)$, but these functions are not square integrable with respect to the Haar measure on $SL(2,\C)$. 
Rather, they span a generalized linear subspace, $\cal K$. 
Furthermore, they satisfy a kind of analyticity property: they are fully determined by their restriction on $SU(2)$. Hence the space $\cal K$, which does not carry an $SL(2,\C)$-covariant scalar product, is instead isomorphic to the Hilbert space of the $SU(2)$ spin networks, ${\cal H}_{SU(2)}$.

This observation clarifies how the  \slc-covariant dynamics provides amplitudes for the canonical theory, based on $SU(2)$. But it also provides a way to give a Lorentz-covariant description to the canonical states. 
In fact, the isomorphism between ${\cal H}_{SU(2)}$ and $\cal K$
equips boundary states with natural covariance properties:  conventional $SU(2)$ spin networks can be represented as functions on $SL(2,\C)$, in a form where their transformation properties under a local Lorentz transformation are manifest, providing an elegant answer to the question we have started from, and restoring manifest Lorentz covariance in canonical quantum gravity.

The tools which make this link possible are the ``projected'' spin networks introduced by Livine \cite{Livine:2002ak} and developed by Alexandrov and Livine \cite{Alexandrov:2002br,Alexandrov:2002xc,Alexandrov:2010pg}.  In particular, Alexandrov has extensively developed a manifestly Lorentz-covariant  spin networks  formalism \cite{Alexandrov:2007pq,Alexandrov:2008da,Alexandrov:2010pg}. Here we focus on aspects and results of this framework that are of direct value for LQG, disentangling them from Alexandrov's attempts to find alternative models.  In \cite{Dupuis:2010jn}, Dupuis and Livine study a map $f$ that sends a $SU(2)$ spin networks into (a certain class of) projected spin networks. The space $\cal K$ defined by the LQG spinfoam amplitudes satisfies the simplicity constraints and is in the image of this map \cite{Alexandrov:2010pg,Dupuis:2010jn,Conrady:2008ea}. In a paper appearing in parallel with this one \cite{Wieland:2010ec}, Wieland gives another direct derivation of the fact that the space $\cal K$ can also be obtained directly from a canonical quantization of general relativity, by using the original self-dual Ashtekar connection as a variable in the Holst action with real Barbero-Immirzi parameter.

The fact that $SL(2,\C)$ functions describe states of canonical LQG, but there is no $SL(2,\C)$-covariant scalar product on the space where they live, is reminiscent of the Gupta-Bleuler formalism \cite{Gupta:1950fk,Bleuler:1950uq}, where the two physical photons can be described in a Lorentz-covariant language, but without positive-definite  Lorentz-covariant scalar product.  The fact that $\cal K$ is not a proper subspace of ${\cal H}_{SL(2,\C)}$ is also reminiscent of loop cosmology, where the state space is taken to be defined by a Bohr compactification of the real line \cite{Ashtekar:2003hd}.  Functions in $\cal K$ are of the same kind: discrete linear combinations of distributions. 

Altogether, these observations show that LQG admits a manifestly Lorentz-covariant formulation, and behaves under Lorentz transformations as expected from classical GR. Like classical GR, the theory is invariant under local Lorentz transformation in the bulk and is covariant under local Lorentz transformation in the boundary.

\section{Dupuis-Livine map} 

Let $\psi(h)$ be a function on $SU(2)$. 
Following Dupuis and Livine \cite{Dupuis:2010jn}, consider a map $f: \psi\mapsto \tilde\psi$ from functions on $SU(2)$ to functions on \slc, as the integral transform
\be
 \tilde\psi(g)=\int_{SU(2)} \hspace{-1em} dh\  K(g,h) \ 
  \psi(h), \hspace{2em} g\in SL(2,\C)
   \label{pf}
\ee
defined by the kernel 
\be
 K(g,h) = \sum_j d_j^2 \int_{SU(2)}\hspace{-1em} dk\ \chi^{p(j),j}(gk)\  \chi^j(kh)  .
   \label{kernel}
\ee
Here $j\!\in\! \N/2$, $d_j=2j+1$, $\chi^j(h)$ is the spin-$j$ $SU(2)$ character and  
$\chi^{p,k}(g)$ is the \slc \ character in the $(p,k)$ representation.\footnote{\slc \  unitary representation in the principle series are determined by the two quantum numbers, $p\in \R$ and $k\in \N/2$, of the two Casimirs $C_1$ and $C_2$, 
\begin{eqnarray}
C_1&\equiv& (1/2)J^{IJ}J_{IJ}= |\vec L|^2 - |\vec K|^2 =p^2-k^2,\\
C_2&\equiv& (1/4) \epsilon^{IJKL} J_{IJ}J_{IJ}= 2 \vec K\cdot \vec L = 2pk, 
\end{eqnarray}
where $J_{IK}, {\scriptstyle I,K}=0,...,3$ are the generators of \slc . Here
$\vec L$ are the generators of $SU(2)\subset SL(2,\C)$ and $\vec K$ the generators of the corresponding boosts. That is, letting $i,j,k=1,2,3$,
\be
L^i=-\frac12 \epsilon^i{}_{jk}J^{jk},\ \ \ \ 
K^i=J^{0i}. 
\ee} 
Finally, $p(j)$ is the assignment of a positive real number $p$ for each $j$, which we call the ``degree" of the map. 

It is not hard to show that 
\be
 (f \psi)\big|_{{}_{SU(2)}}=\psi.
\ee
Therefore the image of $f$ is formed by a linear subspace of the space of functions on \slc, denote it  ${\cal K}$, characterized by the property
\be\label{pf1}
  \tilde\psi(g)= \int_{SU(2)}  dh\  K(g, h)\  \tilde\psi(h). 
\ee 
We call the functions satisfying this property ``projected" functions of degree $p(j)$. 

Notice that these functions satisfy a sort of analyticity property: \emph{they are determined by their restriction on $SU(2)$}.  The space of these functions is therefore linearly isomorphic to a space of functions on $SU(2)$. 
If we define the components of  $\psi$ on the Peter-Weyl basis, 
\be
\psi_{jmn}= \int_{SU(2)} dh\, \overline{D^j_{mn}(h)}\, \psi(h),
   \label{pw}
\ee
where $D^j_{mn}(h)$ are the Wigner matrices; then, \eqref{pf} can be rewritten
\be
    \tilde\psi(g)= \sum_{jmn} d_j \, \psi_{jmn} \, D^{p(j),j}_{jm,jn}(g),
    \label{pf2}
\ee
where $D^{p,k}_{jm,j'n}$ are the matrix elements of the $(p,k)$ representation in the $|(p,k);j,m\rangle$ basis that diagonalizes $L^2$ and $L_z$ of the canonical $SU(2)$ subgroup.  

An important aspect of these functions is that the space ${\cal K}$ spanned is \emph{not} a proper subspace of $L_2[SL(2,\C)]$. This can be better seen by introducing in  $L_2[SL(2,\C)]$ the generalized basis $|p,k,j,m,j',m'\rangle$, defined by 
\be
\langle g |p,k,j,m,j',m'\rangle = D^{p,k}_{jm,j'm'}(g).
\ee
The basis vector are orthogonal,
\begin{eqnarray}
&&\langle \tilde p,\tilde k,\tilde j,\tilde m,\tilde j',\tilde m' |p,k,j,m,j',m'\rangle 
\label{scpr} \\ \nonumber &&\hspace{3em}
=\frac{\delta(p-p')}{(p^2\!+\!k^2)} \delta_{k \tilde k} \delta_{j \tilde j}  \delta_{j' \tilde j'}  \delta_{m \tilde m}  
\delta_{m' \tilde m'}.
\end{eqnarray}
The key point is that $p$ is a continuous label. Therefore normalizable states can be obtained only by \emph{integrating} in $p$,
\be\nonumber
|\psi\rangle=\sum_{k\ldots m'} \int dp  
(p^2\!+\!k^2)
  \psi_{kjmj'm'}(p) |p,k,j,m,j',m'\rangle
\ee
with $\psi_{kjmj'm'}(p)$ square integrable in $p$.
But for $|\psi\rangle$ to be in ${\cal K}$ it must be of the form 
\be
\psi_{kjmj'm'}(p)  = \frac{\delta\big(p-p(k)\big)}{(p^2+k^2)} \delta_{jk} \delta_{j'k}\, \psi_{jmm'}
\ee
which is not square integrable in $p$. In other words, the fixed relation between the continuous variable $p$ and the discrete variable $k$ forces the states in ${\cal K}$ to be a discrete linear combinations of distributions.  

It follows that the $SL(2,\C)$ scalar product is not well defined on ${\cal K}$. Instead, a scalar product is naturally defined by the linear isomorphism between ${\cal K}$ and a space of functions on $SU(2)$. This amounts essentially in replacing the Dirac delta in \eqref{scpr} with a Kroneker delta (and adjusting the measure factor). That is, since 
\be\nonumber
   f |j,m,m'\rangle =  |p(j),j,j,m,j,m'\rangle; 
\ee
and in $L_2[SU(2)]$
\be\nonumber
\langle \tilde j,\tilde m,\tilde m' |j,m,m'\rangle =\frac{\delta_{j \tilde j}}{d_j}
 \delta_{m \tilde m}  \delta_{m' \tilde m'},
\ee
we can define on ${\cal K}$ the well-behaved scalar product
\be\label{scp}
\langle p(j),j,\tilde j,\tilde m,\tilde j,\tilde m' |p(j),j,j,m,j,m'\rangle =\frac{\delta_{j \tilde j} }{d_j}
\delta_{m \tilde m} \delta_{m' \tilde m'},
\ee
instead of the diverging $SL[2,\C]$ one \eqref{scpr}.\footnote{The reduction of $p$ to a discrete label echos the Bohr compactification of the real line used in loop cosmology  \cite{Ashtekar:2003hd}.} 

\section{Fixing the degree}

All that we said above is valid for an arbitrary degree $p(j)$ of the Dupuis-Livine map. 
Let us now select, once and for all, the degree to be 
\be
p(j)=\gamma j
\ee
where $\gamma$ is a positive real parameter. We still refer to ${\cal K}$ as the space of projected functions with this degree, which is now spanned by $D^{\gamma j,j}_{jm,jn}(g)$. 

The interest in this space ${\cal K}$ comes from the fact that this space implements the linear simplicity constraints of general relativity \cite{Engle:2007uq,Rovelli:2010wq,Ding:2010ye}. At the classical level, these can be written in the time gauge as
\be
  \vec K + \gamma\vec L =0.
   \label{sc}
\ee
From this expression we can extract the following gauge-invariant part, 
\be\label{sc2}
2\gamma C_1-(\gamma^2-1)C_2=0.
\ee
One can then show \cite{Engle:2007uq,Rovelli:2010wq,Ding:2010ye} 
that for all $\tilde\psi$ and $\tilde\psi'$ belonging to ${\cal K}$,
the condition \eqref{sc2} is satisfied strongly,
\be\label{CC}
\big(2\gamma C_1-(\gamma^2-1)C_2\big) |\tilde\psi\rangle=0,
\ee
and \eqref{sc} weakly,
\be
  \langle \tilde\psi| \vec K + \gamma\vec L |\tilde \psi'\rangle=0
  \label{KL}
\ee
in the limit $j\mapsto\infty$.\footnote{It is also possible to satisfy the condition \eqref{KL} for all spins \cite{Ding:2010ye,Alexandrov:2010pg}, if one chooses $p=\gamma(j+1)$, but this would violate the cylindrical consistency of the spin foam amplitude \cite{Rovelli:2010qx,Magliaro:2010ih}.}
Here the scalar product is the one determined by the \slc\ Haar measure.  

That is, ${\cal K}$ is a linear subspace of $L_2[\slcm]$ where equations \eqref{CC} and \eqref{KL} hold.
The first condition imposes $p=\gamma k$, and the second one fixes $k=j$, the minimal spin of the canonical $SU(2)$ subgroup.
These are the linear simplicity constraints used in the new spin foam models for quantum general relativity \cite{Engle:2007uq,Rovelli:2010wq}.

At the classical level, these constraints guarantee that the full covariant dynamics can be encoded in the $SU(2)$ Ashtekar-Barbero connection \cite{Ashtekar87,Barbero:1995ud},
$A^i=\omega^i+\gamma\omega^{0i}$, where $\omega^i=-\frac12\epsilon^i_{jk}\omega^{jk}$ and $\omega^{IJ}$ is the full \slc\ connection. Indeed,  let
\be
\omega=\omega^{IK}J_{IK}=-\omega^{0i}K_i+\omega^i L_i
\ee
be  
an \slc\ algebra element. If the condition \eqref{sc} between generators holds, we have
\be
\omega|_{{\cal K}} =(\omega^i +\gamma\omega^{0i})L_i \equiv A^i L_i.
\label{a}
\ee

At the quantum level, the correspondence \eqref{a} is lost since the connection is not a well-defined operator by itself. Only the holonomy, namely the exponential of the connection along a finite path, is. Therefore \eqref{a} is replaced by the relation between the $\slc$ holonomy $g$ and the $SU(2)$ holonomy $h$ induced by \eqref{pf},
\be
g|_{{\cal K}} = D^{\gamma j,j}_{jm,jn}(g) = \int_{SU(2)} \hspace{-1em} dh\  K(g,h) \ D^j_{mn}(h).
\ee
This relation guarantees that the \slc\ holonomy is fully determined by its restriction to $SU(2)$.\footnote{For a discussion on the splitting $A^i=\omega^i+\gamma\omega^{0i}$ at the discrete level, see also \cite{Freidel:2010aq,Rovelli:2010km}.}

After these preliminaries, we are now ready to get to our main subject.

\section{Transition amplitudes}

Following \cite{Geloun:2010vj,Rovelli:2010vv}, the transition amplitudes of LQG can be written in the form 
\begin{eqnarray}
&&Z_{\cal C}(h_l)= \int_{SL(2,\C)} 
dg_{ev}\int_{SU(2)} 
dh_{e\!f}\;
\label{int1}\\\nonumber
&&\hspace{1em} \times \sum_{{j_{\!{}_f}}}  \prod_{f} d_{j_{\!{}_f}}\;
\chi^{\scriptscriptstyle\gamma {j_{\!{}_f}},{j_{\!{}_f}}}\!\left(\!\prod_{e\in\partial f}g_{e\!f}^{\epsilon_{e\!f}}\!\right) \prod_{e\in\partial f}\chi^{j_{\!{}_f}}\!(h_{e\!f})
\end{eqnarray}
Here $\cal C$ is a combinatorial two-complex with vertices $v$, edges $e$ and faces $f$, bounded by a graph $\Gamma=\partial {\cal C}$ with nodes $n$ and links $l$ (See Figure 1). Inside the \slc\ characters, $\epsilon_{ef}$ is a sign, and the quantity $g_{ef}$ is defined by 
\be
g_{ef}= \left\{\begin{array}{ll}
 g_{es_e} h_{e\!f}g^{-1}_{et_e}& {\rm for\ internal\ edges},  \\
h_l\!\in\! SU(2)&  {\rm for\ boundary\ edges}.
\end{array}\right.
\label{gef}
\ee 
Here $s_e$ and $t_e$ are respectively the source and target vertices of the edge $e$.
For the rest of the definition, see  \cite{Rovelli:2010vv}. 
\begin{figure}[t]
\centerline{
\includegraphics[scale=0.15]{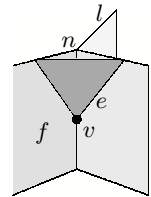}
\hspace{-5.6em}
\begin{picture}(40,30)
\put(20,23){\small $v$}
\put(25,34){\small $e$}
\put(2,22){\small $f$}
\put(12,57){\small $n$}
\put(25,67){\small $l$}
\end{picture}\hspace{5em}
}\caption{A two-complex with one bulk vertex.}\vspace{-1em}
\label{13}
\end{figure}

Notice that in this definition the $SU(2)$ elements $h_l$ only enter inside the \slc\ characters. It follows that $Z_{\cal C}(h_l)$ is in fact the restriction to $SU(2)$ of the function on \slc\ defined by 
\be
\tilde Z_{\cal C}(g_l)= {\rm same\ as\ (\ref{int1},\ref{gef})\ with\ } h_l\ {\rm replaced\ by}\ g_l.
\ee

We now begin elucidating the properties of these transition amplitudes, at the light of the mathematics discussed in the two preceding sections.  We draw largely from the work of Alexandrov and Livine \cite{Livine:2002ak,Alexandrov:2002br,Alexandrov:2002xc,Alexandrov:2007pq,Alexandrov:2008da,Alexandrov:2010pg,Dupuis:2010jn}. See also Conrady and Freidel in \cite{Conrady:2008ea}. 

The first important result is the following. 
\begin{theorem}
$\tilde Z_{\cal C}(g_l)$ is a projected function with degree $p(j)=\gamma j$ in each of its entries.  Equivalently:
\be
        (\otimes_l f) Z_{\cal C}= \tilde Z_{\cal C}.
\ee
\end{theorem}
This can be shown by an explicit computation, inserting \eqref{int1} into the definition of projected functions \eqref{pf}. The computation is straightforward, although somewhat tedious, and we omit the details. The key reason for which the result holds is that the $h_l$ are directly sandwiched between the variables $h_{ef}$ and  $h_{e'f}$, where $f$ is the face bounding $l$ and $e$ and $e'$ are the edges bounding the two nodes that bound $l$. (Recall that there is no $g$ integration at the boundary nodes.) The integrations over $h_{ef}$ and  $h_{e'f}$, amount to projections on the $j=k$ $SU(2)$ subspace of the \slc\  representation, thus trivializing the integrals in the definition of $f$.

As a consequence of this theorem, the $SU(2)$-invariant boundary space ${\cal H}_{LQG}$ is naturally mapped into the $SU(2)$-invariant tensor product of spaces ${\cal K}$ at each link:
\begin{eqnarray} \nonumber
&& \psi_{[\Gamma,j_l,i_n]}(h_l) = \otimes_l D^{j_l}(h_l) \otimes i_n \\ \nonumber
&& \mapsto \tilde\psi_{[\Gamma,j_l,i_n]}(g_l) = \otimes_l D^{\gamma j_l,j_l}(g_l) \otimes i_n,
\end{eqnarray}
where $i_n$ are $SU(2)$ intertwiners, and contraction over the magnetic indices is tacitly assumed.

\section{Lorentz covariance}

The point of our main interest is the restoration of manifest local Lorentz covariance of the boundary space.
In classical general relativity, spacetime is assumed to be a pseudo-Riemannian manifold, and a Minkowksi metric is defined on the tangent space of each spacetime point. The Lorentz group $SO(3,1)$ is the symmetry group of Minkowski space, and in this sense the dynamics of general relativity is locally Lorentz invariant. The symmetry is manifest in the tetrad formalism, where the gravitational field is described by a one form with values in Minkowski space, and the GR action is invariant under local Lorentz transformations in this space.  

In the physical theory, then, \slc\  represents the covering group of the part connected to the identity, $SO_0(3,1)$, of the group of the local Lorentz transformation, while $SU(2)$ represents the covering group of the $SO(3)$ group of rotations of the physical space defined by a certain Lorentz frame. If we view $SU(2)$ and $\slcm$ as groups of matrices, then $SU(2)$ is naturally a subgroup of \slc. 
Let us call $x_o$ this canonical embedding: $x_o(h)=h$.
From the point of view of physics, on the other hand, 
there is no preferred embedding of  the abstract group $SU(2)$ into the abstract group \slc. 
If we select a unit timelike vector $x$ on the hyperboloid ${\mathbbm H}^3$, namely a local Lorentz frame, then the subgroup $SO(3)_x\subset SO_0(3,1)$ that leaves $x$ invariant defines the group of rotations of physical space.   
The canonical embedding as matrix groups, $x_o$, simply corresponds to a special choice of vector.
Hence, there is an ${\mathbbm H}^3$-worth of isomorphisms, which we also denote by $x$, of $SU(2)$ into \slc: one per each possible state of motion of an observer at a spacetime point. 
Calling $SU(2)_x$ the image of $x$, we have the embeddings
\be 
h_x=x(h)\in SU(2)_x\subset \slcm. 
\ee
Fixing a reference vector, say $x_o=(1,0,0,0)$ in some coordinates, each $x$ defines a Lorentz transformation $\Lambda_x$ which is a pure boost and sends $x_o$ to $x$. Clearly, 

\be
     h_x=\Lambda_x h \Lambda_x^{-1}.
\ee
  
Furthermore, one can also consider more general embeddings $h\to h_{xx'}$, of the form
\be
     h_{xx'}=\Lambda_x h \Lambda_{x'}^{-1},
\ee
where $x$ and $x'$ may be different. Such embeddings are motivated if we view $h$ as the parallel trasport between two points, and $\Lambda_x$, $\Lambda_{x'}$ as gauge transformations.
The image $SU(2)_{xx'}$ of this map is a subgroup only if $x=x'$.

Given one of these embeddings from $SU(2)$ into \slc, we have immediately a map from functions on \slc\ to functions  on $SU(2)$, simply obtained restricting the former to the image of the map.

The Dupuis-Livine map is also defined for such arbitrary embeddings $h\mapsto h_{xx'}$. We have
\be
f_{xx'}: \psi\mapsto \tilde\psi_{xx'},
\ee
\vspace{-1em}with
\be
 \tilde\psi_{xx'}(g)=\int_{SU(2)}dh\  K_{xx'}(g,h) \ 
  \psi(h),
 \label{pff}
\ee
and the kernel given by 
\be
 K_{xx'}(g,h) = \sum_j d_j^2 \int_{SU(2)} dk\ \chi^{\gamma j,j}(gk_{xx'})\  \chi^j(kh)  .
\nonumber     \label{pf5}
   \ee
   Here we have already fixed the degree of the map that is relevant for quantum general relativity.

As before, one can easily check the projection property
\be 
 (f_{xx'} \psi)\big|_{{}_{SU(2)_{xx'}}}=\psi.
\ee
The image of $f_{xx'}$ is formed by a linear subspace ${\cal K}_{xx'}$ 
of the space of functions on \slc\ characterized by the property
\be
  \tilde\psi_{xx'}(g)= \int_{SU(2)}  dh\  K_{xx'}(g, h)\  \tilde\psi_{xx'}(h_{x'x}). 
\ee 
These functions are determined by their restriction on $SU(2)_{x'x}$, and the 
space of these functions is still isomorphic to $L_2[SU(2)]$.  They have the form
\be
    \tilde\psi_{xx'}(g)= \sum_{jmn} \ d_j \psi_{jmn}\  D^{\gamma j,j}_{jm,jn}(\Lambda^{-1}_{x'} g \Lambda_x).\ 
    \label{pf24}
\ee

Notice now that the transition amplitudes \eqref{int1} are defined in terms of the embedding $x_o$ of $SU(2)$ into \slc. But we have observed that from the point of view of physics the two groups are abstract groups and there is no preferred embedding.  Disregarding this fact leads to a formulation of the theory in which a certain Lorentz gauge has been chosen at each point.  Let us instead look for a formulation where the covariance under the choice of this gauge is left explicit. For this, pick a unit timelike vector $x_e$ at each edge of the 2-complex, and generalize the definition of the transition amplitudes \eqref{int1} to the form 
\begin{eqnarray}
&&\tilde Z_{{\cal C},x_e}(g_l)= {\rm same\ as\ (\ref{int1})\ with\ } g_{ef}\ {\rm given\ by}\nonumber \\[2mm]
&& g_{ef}= \left\{\begin{array}{ll}
g_{es_e} (h_{e\!f})_{x_e}g^{-1}_{et_e}& {\rm for\ internal\ edges},  \\
(h_l)_{x_{s(l)}x_{t(l)}}&  {\rm for\ boundary\ edges}.
\end{array}\right.
\nonumber
\label{gef2}
\end{eqnarray}
It is then immediate to derive the second key result.
\begin{theorem}
$\tilde Z_{{\cal C},x_e}(g_l)$ is independent from all $x_e$ where $e$ is a bulk edge.
\end{theorem}
This follows trivially from the fact that all $\Lambda_x$ group elements can be reabsorbed into the \slc\ integrations. This is a simple but important result, because it shows explicitly that the dynamics of the theory is Lorentz invariant in the bulk. 

Hence, the transition amplitudes depend only on the $x$'s on the boundary edges.  Since there is one of these per each node $n$ of the boundary graph, it is convenient to write $\tilde Z_{{\cal C},x_e}(g_l)$ in the form $\tilde Z_{{\cal C},x_n}(g_l)$.  

Finally, we can study the covariance properties of the amplitude. 
\begin{theorem}
Under a local Lorentz transformations on the boundary, the transition amplitudes transform in the following way
\be 
\tilde Z_{{\cal C},\Lambda_n x_n}(g_l)= \tilde Z_{{\cal C},x_n}(\Lambda_{s_l} g_l \Lambda_{t_l}),
\ee
where $s_l$ and $t_l$ are the source and target of the link $l$.
\end{theorem}
The result is a direct consequence of the transformation properties of projected spin networks.
This is the correct covariance property of the \slc \ holonomy under gauge transformations. 

\section{Conclusion}

We have studied the covariance properties of the LQG transition amplitudes under local Lorentz transformations.  We have shown that the amplitudes are invariant under local gauge transformations in the bulk (Theorem 2). On the boundary, there exists a manifestly Lorentz covariant formalism, given by a certain class of  ``projected spin networks". 

The Dupuis-Livine map that sends LQG boundary states into projected spin networks trivializes for the transition amplitudes, in the sense that these amplitudes are in fact naturally defined as \slc\ functions that satisfy the condition defining the relevant class of projected spin networks (Theorem 1).  It follows immediately that the transition amplitudes transform properly under local gauge transformations on the boundary (Theorem 3).

The restriction to a special class of projected spin networks is motivated by the simplicity constraints. 
Remarkably, the same space $\cal K$ can be obtained from a canonical quantization, as discussed by Wieland in a paper that is appearing in parallel with this one \cite{Wieland:2010ec}. Wieland's results can be interpreted as follows. Start from the Holst action
\be
S[e,\omega]=\int [(e\wedge e)^*+\frac1\gamma(e\wedge e)]\wedge F[\omega]
\ee
and fix the time gauge obtained demanding that the restriction of $e$ to the boundary satisfies $ne=0$, where $n$ is a scalar with values in Minkowski space. The momentum conjugate to $\omega$ is immediately read out of the action:
\be
\pi=(e\wedge e)^*+\frac1\gamma(e\wedge e).
\ee
In the time gauge, it satisfies
\be
K:=n\pi=(e\wedge e)^*,\hspace{1em}L:=-n\pi^*=-\frac1\gamma(e\wedge e)^*,
\ee
where $K$ and $L$ are its electric and magnetic components in the time gauge. The linear simplicty constraint \eqref{sc} follows immediately. Notice that  \slc\ has a natural complex structure and we can define the complex variables $\Pi=K+iL$ and $\overline \Pi=K-iL$. Then \eqref{sc} can be interpreted as a reality condition.   If we quantize the theory in terms of \slc\ cylindrical functions, then $\pi$ becomes the \slc\ generator, and we can impose  \eqref{sc} simply by chosing a scalar product whith respect to which this reality condition is realized. On the space of functions on  \slc, the scalar product \eqref{scp}
is precisely a solution to this problem, and defines ${\cal K}$. Thus the same ${\cal K}$ we have derived here from the spinfoam amplitudes can also be obtained via straightforward canonical quantization of the Holst action, using the old idea of implementing the reality conditions as the conditions that determines the scalar product (see e.g. \cite{Rovelli:1991zi}).  

The possibility of a Lorenz covariant formulations of spin networks has been extensively studied by Alexandrov in \cite{Alexandrov:2007pq,Alexandrov:2008da,Alexandrov:2010pg}, where several of of the results presented here can be already found, framed in a different context. See also \cite{Cianfrani:2008zv}.

In summary, the dynamical diffeomorphism invariant quantum field theory defined by the transition amplitudes \eqref{int1} appears to be fully consistent with the local Lorentz invariance of general relativity.  

\centerline{---}

We thank Thomas Krajewski and Abhay Ashtekar for very useful conversations, Wolfgang Wieland for sharing his results with us before publication, and Sergei Alexandrov for an extensive mail exchange. 


\end{document}